\documentclass{IEEEtran4PSCC}
\usepackage{subcaption}
\ifCLASSINFOpdf
   \usepackage[pdftex]{graphicx}
\else
   \usepackage[dvips]{graphicx}
\fi
\usepackage[cmex10]{amsmath}
\usepackage{amsfonts}
\usepackage{hyperref}
\usepackage{xcolor}
\usepackage[linesnumbered,ruled]{algorithm2e}

\makeatletter
\let\old@ps@headings\ps@headings
\let\old@ps@IEEEtitlepagestyle\ps@IEEEtitlepagestyle
\def\psccfooter#1{%
    \def\ps@headings{%
        \old@ps@headings%
        \def\@oddfoot{\strut\hfill#1\hfill\strut}%
        \def\@evenfoot{\strut\hfill#1\hfill\strut}%
    }%
    \def\ps@IEEEtitlepagestyle{%
        \old@ps@IEEEtitlepagestyle%
        \def\@oddfoot{\strut\hfill#1\hfill\strut}%
        \def\@evenfoot{\strut\hfill#1\hfill\strut}%
    }%
    \ps@headings%
}
\makeatother

\psccfooter{%
        \parbox{\textwidth}{\hrulefill \\ \small{} \hfill \hfill \small{}}%
}

\begin{document}
%
\title{GridLearn: Multiagent Reinforcement Learning for Grid-Aware Building Energy Management}

\author{
\IEEEauthorblockN{Aisling Pigott, Constance Crozier, Kyri Baker}
\IEEEauthorblockA{Dept of Civil, Environmental, Architectural Engineering \\
University of Colorado Boulder\\
Boulder, CO, USA\\
\{aisling.pigott, constance.crozier, kyri.baker\}@colorado.edu}
\and
\IEEEauthorblockN{Zoltan Nagy}
\IEEEauthorblockA{Dept of Civil, Architectural, Environmental Engineering \\
The University of Texas at Austin\\
Austin, TX, USA\\
nagy@utexas.edu}
}

\maketitle

\begin{abstract}
Increasing amounts of distributed generation in distribution networks can provide both challenges and opportunities for voltage regulation across the network. Intelligent control of smart inverters and other smart building energy management systems can be leveraged to alleviate these issues. GridLearn is a multiagent reinforcement learning platform that incorporates both building energy models and power flow models to achieve grid level goals, by controlling behind-the-meter resources. This study demonstrates how multi-agent reinforcement learning can preserve building owner privacy and comfort while pursuing grid-level objectives. Building upon the CityLearn framework which considers RL for building-level goals, this work expands the framework to a network setting where grid-level goals are additionally considered. As a case study, we consider voltage regulation on the IEEE-33 bus network using controllable building loads, energy storage, and smart inverters. The results show that the RL agents nominally reduce instances of undervoltages and reduce instances of overvoltages by 34\%.

\end{abstract}

\thanksto{\noindent This work utilized resources from the University of Colorado Boulder Research Computing Group, which is supported by the National Science Foundation (awards ACI-1532235 and ACI-1532236), the University of Colorado Boulder, and Colorado State University.}

\section{Introduction}

Towards pursuing a reliable and cleaner energy system, solar photovoltaic (PV) generation is being increasingly connected into electricity distribution networks. In 2019, there was estimated to be 23.2 GW of small-scale PV in the US~\cite{IEAsolar}, and this number is rapidly increasing each year. Distributed generation is an effective method for reducing carbon emissions from the electricity sector. However, the integration of PV into a distribution network can cause voltage issues~\cite{Dubey2017}, so there is a limit to the amount of solar that can be installed without intervention~\cite{8973688,Ding16}.

The cost of upgrading distribution networks to increase PV penetrations is high, however the use of smart inverters and intelligently shifting building loads to overlap with solar production can avoid upgrades in many cases~\cite{osti_1432760}. Smart inverters can vary the ratio of real and reactive power that is exported to the network, thus providing voltage support (known as volt-VAR control). Additionally, it has been suggested that energy storage and demand response could be used for voltage regulation~\cite{6465666}. Therefore, it is likely that a building with flexible loads and/or smart inverter-connected PV could be exploited, to lessen voltage issues in distribution networks with a high PV penetration. 

Various methods have been proposed for controlling assets to achieve voltage regulation in distribution networks. In \cite{Nasiri2016} two centralized control methods are proposed. However, centralized control algorithms typically require a full distribution network model, while utilities may not have accurate network models. Centralized control schemes also suffer from issues with privacy and response time~\cite{5759191}. 

Decentralized methods have also been investigated for use in voltage regulation, where the control algorithm is applied individually by each inverter/household. Standard volt-VAR controls determine a policy for reactive power injection based on the voltage at the point of connection. These do not require a full network model, however they can sometimes cause unintended oscillatory behaviour~\cite{Hoke15}. More sophisticated decentralized methods, such as using particle swarm optimization~\cite{Zafar2016}, or a consensus negotiation~\cite{Wang2019} have also been proposed. However these, in addition to many decentralized methods that improve upon traditional volt-VAR, require an accurate model of the distribution network~\cite{Baker18_Network}. 

Reinforcement learning (RL) allows the control agents to learn from experience on the network, meaning a network model is not required, and agents can dynamically adapt to a changing environment. This means that, unlike previously discussed approaches, the algorithm will adapt when other buildings are taken on/offline or modified. Centralized RL has been previously investigated for voltage regulation~\cite{Cao2020,Ying21}. However, both only consider optimization of reactive power components; in \cite{Cao2020} demand is considered to be constant, while in \cite{Ying21} load is inflexible. In reality, the flexibility provided by appliances or battery energy storage can also aid voltage regulation. It is also important to include varying electricity demand because voltage issues tend to occur at specific times of day, rather than constantly. Finally, these methods also suffer from the aforementioned drawbacks of centralized control algorithms. 

Towards intelligent load shifting and demand response coupled with building-level energy goals, CityLearn is an OpenAI-Gym environment that allows for the implementation of RL strategies to achieve building and community level goals \cite{citylearn19,citylearn20}. However, objectives such as grid-level voltage regulation cannot currently be considered in CityLearn, as no power network model is included in the environment. Without the consideration of grid-level objectives and constraints, buildings are subsequently limited in the amount of demand response they can provide - constrained by grid-level voltage, current limits, and more. Shifting building loads across an entire distribution network can also have adverse effects on the grid, such as on voltages, as will be demonstrated in this paper.

To overcome these issues, we develop a model-free multi-agent reinforcement learning (MARL) framework that addresses both building-level and grid-level objectives, building upon the powerful CityLearn framework.  MARL is preferential to a centralized reinforcement learning approach, because it considers privacy constraints of each building owner and can utilize distributed computing. Previous works have used MARL for other building energy management objectives~\cite{Aladdin2020,Vazquez20}. Here we extend this framework to include PV inverters and a voltage regulation objective, which requires power flow analysis to be incorporated into the environment. Additionally, unlike \cite{Vazquez20} we incorporate synchronous action selection, which limits the RL agents' ability to identify the strategies of other agents \cite{Shoham2007}. Given that all buildings are working towards a communal goal, cooperation between buildings should be incentivized. A lack of coordination could result in over-correcting for whatever issue they may face. However, we have chosen to analyze the worst case scenario, in which building operators cannot coordinate with other buildings in advance.  

Therefore, the contributions of this paper can be summarized as follows. First, we develop a flexible framework (GridLearn) for training RL agents to satisfy grid-level goals while also considering building-level goals and demand response capabilities, building on the existing open-source CityLearn framework. Second, we investigate the use of MARL for voltage regulation, an important emerging issue in high-penetration PV networks, on the IEEE 33-bus network. Third, we validate the use of MARL for building energy management in the extreme case where actions are implemented simultaneously, which can potentially cause oscillatory behavior (e.g., the ``rebound effect'' \cite{Lutolf18}).

The rest of this paper is organized as follows: in Section \ref{sec:rl_env} we provide context for the reinforcement learning algorithms used; in Section \ref{sec:gridlearn} we review the energy models from CityLearn and provide a new power flow model; in Section \ref{sec:sim_results} we provide a brief study of MARL using a voltage deviation objective; finally, in Section \ref{sec:conclusion} we consider other applications for the framework introduced in this paper.

\section{Reinforcement Learning Environment}
\label{sec:rl_env}
In this paper we introduce GridLearn\footnote{https://github.com/apigott/CityLearn/releases/tag/gridlearn-v1.0}, an adaptation of the CityLearn environment~\cite{Vazquez20}. CityLearn is an open source project for MARL using on-site building energy storage to achieve demand response goals. Success is measured by metrics such as flattening load ramp rates, reducing peak electric demand, and minimizing net consumption. In GridLearn, we add voltage regulation metrics by leveraging the Python library pandapower for power flow calculations. To this end, the buildings in GridLearn are equipped with volt-VAR controlled smart inverters in addition to controllable loads for thermal energy storage.

\begin{figure}
    \centering
    \includegraphics[width=\linewidth]{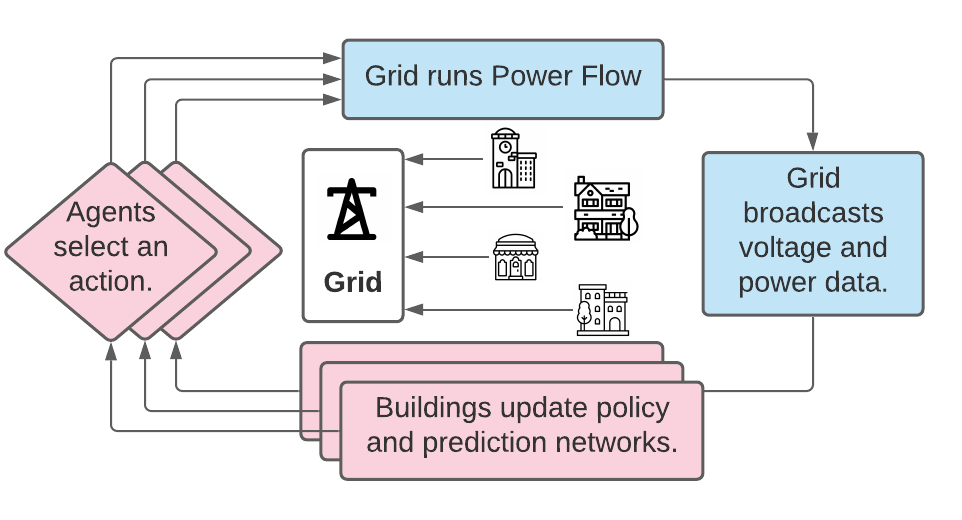}
    \caption{Agent-environment interaction cycle}
    \label{fig:architecture}
\end{figure}

The case study included in this paper is based on the IEEE 33-bus distribution network and includes 6 buildings per bus. These buildings implement RL control as described in Fig \ref{fig:architecture} with 50\% of buildings implementing RL control. The remaining houses implement rule based control (RBC). The use of RBC buildings in addition to RL houses allows for a higher load that causes more undervoltages, while speeding up the simulation time and reducing the required computational power. A lower penetration of RL agents also indicates that community distribution networks can be improved even when the buy-in from building owners is not 100\%.

\subsection{Reinforcement Learning}
In its model-free form, reinforcement learning is a mechanism for online learning of agent-environment interaction. As the RL agent spends more time training, it develops a function or network that characterizes the observed \emph{reward} as a function of the current state and action pair, $(x_t,u_t)$, rather than attempting to predict the next state. RL is therefore particularly useful in cases where the system dynamics would be too hard to model in optimization constraints, as is required in model predictive control. A variation on RL, deep reinforcement learning (DRL), replaces the function approximators with neural networks updated through backpropogation. In this paper we use DRL which is better suited to complex environments.

RL agents are able to provide optimal control policies by extrapolating the current state and action to the future reward it might achieve for the most likely control trajectory. This score is called $q$ and considers the likelihood of choosing an action, and the predicted score achieved by that action:

\begin{equation}
    q(x_t,u_t) = \mathbb{E}[r(x_{t+1}) + \gamma v_\phi(x^{t+1})]
    \label{eq:bellman}
\end{equation}

\noindent Let $v_\phi$ be the network \emph{value}, or expected $q$-value of future states, and the discounting factor, $\gamma$, be a negligible value of 0.99 for continuous control problems.

The $q$ network approximator can be updated periodically through backpropogation to minimize the mean squared error given in \eqref{eq:bellman_error}. Note that a target value, $Q$, can be found after the action is implemented and the environment is updated to the state $x^{t+1}$ (Eq. \eqref{eq:target_q}).

\begin{align}
    \delta(x_t,u_t) &= || Q(x_{t+1}) - q(x_t, u_t) ||^2 
    \label{eq:bellman_error} \\
    Q(x_{t+1}) &= r(x_{t+1}) + \gamma v_\phi(x_{t+1})
    \label{eq:target_q}
\end{align}

Once the RL agent can accurately predict the relationship between action and reward, it can maximize the expected reward by drawing from the policy distribution. Let $\pi_\theta$ be the parameterized probability distribution for selecting an action, with mean and standard deviations that depend on the state. Therefore $u_t$ is selected from the distribution as $u_t \sim \pi_\theta(u | x_t)$.

For complex environments such as power flow simulations, RL is an efficient method for finding an optimal control policy, without modeling the environment or finding convex relaxations. Sutton and Barto \cite{Sutton1998} provides additional reading on convergence guarantees, variations on classic algorithms, and other components to RL not included in this section.

\subsection{Proximal Policy Optimization}
Each individual RL agent works to optimize their own policy. Updating of these RL policies is handled by the Stable Baselines package \cite{stable-baselines} and uses the Proximal Policy Optimization (PPO) algorithm. One of the novel aspects of PPO is the use of clipping to reduce variance in the policy as the agent learns \cite{schulman2017proximal}. 

Vanilla reinforcement learning agents update their policy to maximize the expected value of the reward:
\begin{equation}
    \max_\theta \bigg(\mathbb{E}[\pi_\theta(u_t | x_t) (q(x_t, u_t) - v_\phi(x_t)]\bigg)
    \label{eq:vanilla}
\end{equation}
\noindent where $\pi_\theta$ is the parameterized policy network at time $t$, $q(x_t, u_t)$ is the observed $q$-value of the RL agent, and $v_\phi$ is the parameterized value network. The value approximation can be conceptualized as the expected value of $q$ over all actions in the action space. Therefore Equation \eqref{eq:vanilla} updates the parameterization of the policy to maximize the probability of selecting an action that is \emph{advantageous} (providing a higher reward than could be expected).

To improve upon the vanilla policy optimization, the PPO algorithm represents the change in policy as a time-varying ratio, $r_t$:
\begin{equation}
    r_t = \dfrac{\pi_\theta^{t}(u_t | x_t)}{\pi_\theta^{t-1}(u_t | x_t)}
\end{equation}

\noindent and modifies vanilla policy updates by constraining the policy update. Constraining the policy update was first introduced in the Trust Region Policy Optimization (TRPO) algorithm as explicit constraints on the parameterization $\theta$, but has since been more widely adopted as part of the PPO algorithm. PPO constrains the change in policy as follows:
\begin{align}
    &\max_\theta \left( \min_\theta \big\{ r_t(\theta) (q(x_t, u_t) - v_\phi(x_t)), \hat{r}_t(\theta) \big\} \right) \\
    &\hat{r}_t(\theta) = clip(r_t(\theta), 1 - \epsilon, 1 + \epsilon)
\end{align}
\noindent where $\hat{r}_t$ is a dynamic ratio clipped by a measure of divergence of the new policy from the old one, $\epsilon$.

\subsection{Multi-Agent Reinforcement Learning Framework}

In this framework, each building, which is comprised of up to four energy systems and four control decisions, has their own RL agent. The RL agent uses 18 state space values such as: outdoor air temperature, current state of charge, and time of day to predict an optimized set of actions corresponding to HVAC thermal energy storage, DHW thermal energy storage, PV curtailment, and inverter phase lag. GridLearn allows for the user to specify which of these resources are available to the RL agent.

The implementation of one RL agent with agent-environment interaction is a standard functionality of OpenAI gym \cite{brockman2016}. To add multiple independent agents to the same environment we use a combination of the PettingZoo \cite{terry2020pettingzoo} and SuperSuit \cite{SuperSuit} libraries. PettingZoo is a multi-agent reinforcement learning wrapper that combines multiple agents' actions before passing them to the OpenAI gym environment (which takes just one action argument); SuperSuit provides pre-processing of the environment and allows for agents in the grid environment to have a non-uniform actionspace as dictated by the number of available energy systems in their buildings. 

\section{GridLearn Environment}
\label{sec:gridlearn}
The GridLearn environment provides energy models of many buildings in a mixed use district, connected by a distribution network modeled with AC power flows.

\subsection{Building Models}

The CityLearn environment provides 9 energy models created in EnergyPlus. These buildings represent a combination of office buildings, multifamily residential buildings, restaurants and retail spaces. While the EnergyPlus demand profiles are fixed, each building also has thermal energy storage in the form of indoor air temperature and the hot water tank. The thermal energy storage is tracked in Python and corresponds to a thermal deadband for both thermal energy system. Battery energy storage is also created in the Python environment and can be used to supplement energy imports and exports from the grid. Lastly, the phase lag of the smart inverter is used to scale the active and reactive power supplied by the PV and battery. 

\begin{figure*}[t]
    \centering
    \includegraphics[width=\textwidth]{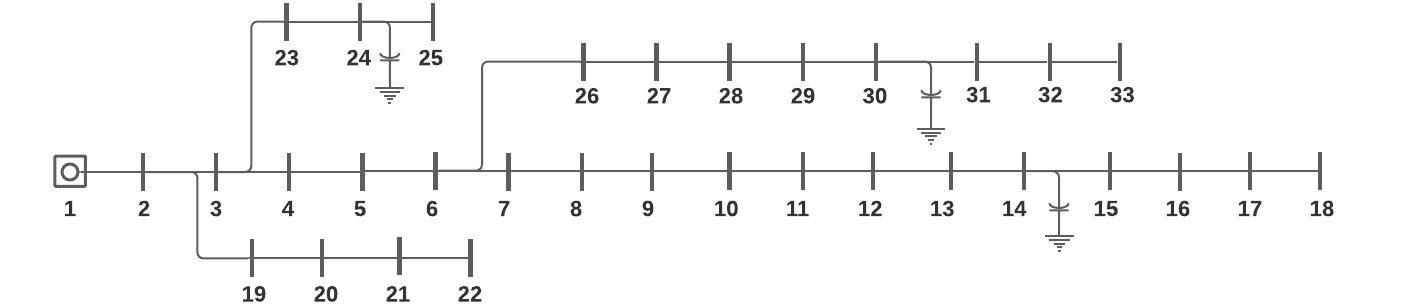}
    \caption{IEEE 33-Bus Distribution Network}
    \label{fig:ieee33}
\end{figure*}

At each timestep, actions that correspond to energy storage are sent to the corresponding device object (HVAC, water heater) to either charge or discharge their storage reserves. Let each energy system be defined by a maximum charge and discharge rate, $\overline{P}$ and $\underbar{$P$}$ respectively, and a maximum storage capacity, $E_{max}$. The energy storage device can then be considered to serve two loads: the first is a mandatory thermal loads calculated by EnergyPlus for occupant comfort, $P_{demand}$, and the second is an external storage device for building energy management, $P_{request}$. Regardless of the charge/discharge signal sent by the RL agent, the thermal load is met first and only then is the storage load considered with the remaining available device power.

When the energy systems receives a charge/discharge signal, it first checks the current state of charge and calculates the maximum amount of energy that can be stored or discharged at that time. The available energy is translated to a constant power signal ($P_{ch}^{avail}$). The energy system then checks the amount of power that must be output to meet the required thermal load, and the maximum charge rate is capped at the remainder of the system's rated power. Similarly the maximum discharge rate is capped at the power requested to meet the thermal load. The process used to determine the resulting power used for to charging/discharging and energy storage system given a control action from the RL agent is described in Algorithm \ref{alg:charge}. Let the input $u$ be a normalized action value $[-1,1]$ where $u < 0$ maps to discharging the device, and $u > 0$ maps to charging the device.

\begin{algorithm}
    \SetKwInOut{Input}{Input}
    \SetKwInOut{Output}{Output}
    \SetKw{Return}{Return}
    \underline{function Charge} $(u)$\;
    \Input{Charge/discharge action, $u$}
    \Output{Electric power consumed}
    $P_{cons}^{elec} = \eta^{-1} P_{demand}$ \\
    $P_{request} = \underbar{$P$} + u (\overline{P} - \underbar{$P$})$ \\
    \eIf{$u\geq 0$}
      {
        $P^{avail}_{ch} \leftarrow (E_{max} - E^t) (\Delta t)^{-1}$ \\
        $P^{avail}_{ch} \leftarrow \min\{P^{avail}_{ch}, \overline{P} - P_{request}\}$ \\
        $P_{stor} = \min\{P_{request}, P^{avail}_{ch}\} $
      }
      {
        $P_{disch}^{avail} \leftarrow E^t  (\Delta t)^{-1} $ \\
        $P_{disch}^{avail} \leftarrow \max\{P^{avail}_{disch}, \overline{P}\}$ \\
        $P_{stor} = \max\{P_{request}, P^{avail}_{disch}\} $
      }
      $E^{t+1} = E^{t} - E^{loss} + P_{stor}\Delta t$ \\
      $P^{elec}_{cons} += \eta^{-1} P_{stor}$ \\
      \Return{$P^{elec}_{cons}$}
    \caption{Algorithm for determining total power consumption of energy storage devices}
    \label{alg:charge}
\end{algorithm}

By using the process described above, the environment (rather than the RL agent) is tasked with validating the control action at any given time. This method ensures that the building environments will serve the required loads, and remain comfortable for the occupant regardless of how well the RL agent performs voltage regulation. Should the RL agent select an action that indicates discharging the storage device when its state of charge is empty or charging when the device is full, the environment simply ignores the charging request.

CityLearn also uses the EnergyPlus weather file to determine PV power production, although the PV inverter in CityLearn is not currently controllable by curtailment or phase lag. Thus, in order to make CityLearn suitable for analyzing the impact of building loads on voltages across the distribution network, the following adaptations were made:

\begin{table}[]
    \centering
    \begin{tabular}{c|c|c}
        \textbf{CityLearn Building Type} & \textbf{Building Description} & \textbf{Num. Buildings} \\
        1 & Medium Office & 4\\
        2 & Fast-food Restaurant & 37\\
        3 & Standalone Retail & 5\\
        4 & Strip Mall & 1\\
        5,6,7,8,9 & Multi-family Residential & 145
    \end{tabular}
    \caption{Building portfolio summary}
    \label{tab:building_portfolio}
\end{table}

\begin{itemize}
    \item Buildings are given a fixed power factor rating of 0.95 lagging.
    \item A variable percentage of buildings are RL-enabled.
    \item Energy demand profiles are replicated to multiple buildings to increase the aggregate load across the network. While CityLearn uses one of each building type, we skew the distribution to favor residential buildings. The building portfolio is summarized in Table \ref{tab:building_portfolio}. 
    \item Energy demand profiles are upsampled from hourly to subhourly intervals. Upsampling is done with linear interpolation for weather data and HVAC, interpolation with added noise for solar irradiance, and by randomly dividing the demand across subhourly intervals for hot water. 
    \item To increase the agency of each RL agent, all buildings are given a PV array.
    \item Smart inverters with phase lag control are added to buildings with DC resources (PV, battery). The addition of smart inverters is necessary to control reactive power demand of the houses. 
\end{itemize}

\subsection{Grid Environment}
Grid models and AC power flows were modeled using pandapower \cite{pandapower2018}. The pandapower library models the loads of the buildings with real and apparent power specifications; the PV arrays (and corresponding inverters) are modelled as PQ controlled generators. Pandapower calculates real and reactive power at each bus, load, and generator along with voltages at each bus. These values can be easily adapted into the state space or reward function. In this study, we used a preconfigured IEEE network model, but the pandapower library has many available distribution networks and supports custom networks.

\section{Simulation Setup}

The study considered in this paper is just one example of how GridLearn can be used for grid level goals. GridLearn is flexible and can adapt to any number of pandapower network configurations, reward functions, and/or agent settings. 

\subsection{Distribution Network}

\begin{table}[]
    \centering
    \begin{tabular}{c|c}
        Total Buildings & 192\\
        RL Enabled Buildings & 96 \\
        PV Penetration [\%] & 100 \\
        Num. Buses & 33 \\
        Time Interval & 15 minutes \\
        ASHRAE Climate Zone(s) & 3A \& 2A
    \end{tabular}
    \caption{Environment Settings}
    \label{tab:env_settings}
\end{table}

The distribution system is modeled using the pandapower library and the IEEE 33-bus network. For each of the 32 distribution nodes 6 buildings are randomly assigned to the node. To help with voltage regulation, a 1.2 MVAR capacitor bank are placed at Bus 14. In the summer months, when undervoltages are more prevalent, three additional capacitor banks are turned on at Buses 14, 24, and 30; they are rated at 0.6 MVAR, 0.6 MVAR, and 1.2 MVAR respectively (Fig \ref{fig:ieee33}). Placing capacitor banks or voltage regulators in distribution grids is common practice in maintaining acceptable average voltages across the network \cite{casillas17}, although as mentioned previously, under high PV penetrations voltage limits can still be violated regularly.

\subsubsection{Reward Function}
In this scenario we propose a reward function based on the voltage deviation from 1 p.u. at the point of common coupling for that building. The reward is calculated independently of all other building owners and ensures complete privacy of all neighboring building owners. Building owners can also measure the voltage independently of the grid operators and with little extra instrumentation. This eliminates the need for peer to peer or client-server communication networks.

In the following reward functions, let $x^t$ represent the measured state values of the building including current weather conditions, voltage data, load data, and state of charge (SOC) of all energy storage devices. Let $v_i$ denote the voltage measured at the nearest bus and $\alpha_i$ be a building-specific weighting factor to approximately normalize the reward function. 

\begin{equation}
    r(x^t) = -(\alpha_i(v_i^t - 1))^2 + 1 
    \label{eq:voltage_reward}
\end{equation}

\subsection{Baseline Scenario}
To reduce the computational complexity of the environment, some of the buildings are controlled with RBC. The RBC agents are based on a diurnal charge and discharge cycle according to the baseline scenario in the CityLearn challenge. This kind of controller would be suited to simple demand response signals such as time-of-use pricing. The RBC agent is used as the baseline comparison tool (100\% RBC penetration) and to supplement RL agents in the RL-scenario ($n$\% RL agents; ($100-n$)\% RBC). 

\section{Simulation Results}
\label{sec:sim_results}
The set of RL agents was experimentally tuned on the environment until they improved on the baseline as measured by voltage deviations from 1 pu. A lower learning rate was found to be particularly helpful in avoiding oscillatory behavior. Additional tuned hyperparameters are provided in Table \ref{tab:rl_settings}, though it should be noted that the environment state, actions, and rewards were also normalized in order to take advantage of most of the default Stable Baselines hyperparameters. 

\begin{table}[b]
    \centering
    \begin{tabular}{c|c}
        Batch size & 64 \\
        Learning rate & $10^{-5}$ \\
        Num. steps per update & 256 \\
        KL clipping value ($\epsilon$) & 0.2 \\
        Num. steps total & $7\times10^4$
    \end{tabular}
    \caption{RL Agent Settings}
    \label{tab:rl_settings}
\end{table}

\begin{figure}
    \centering
    \includegraphics[width=\linewidth]{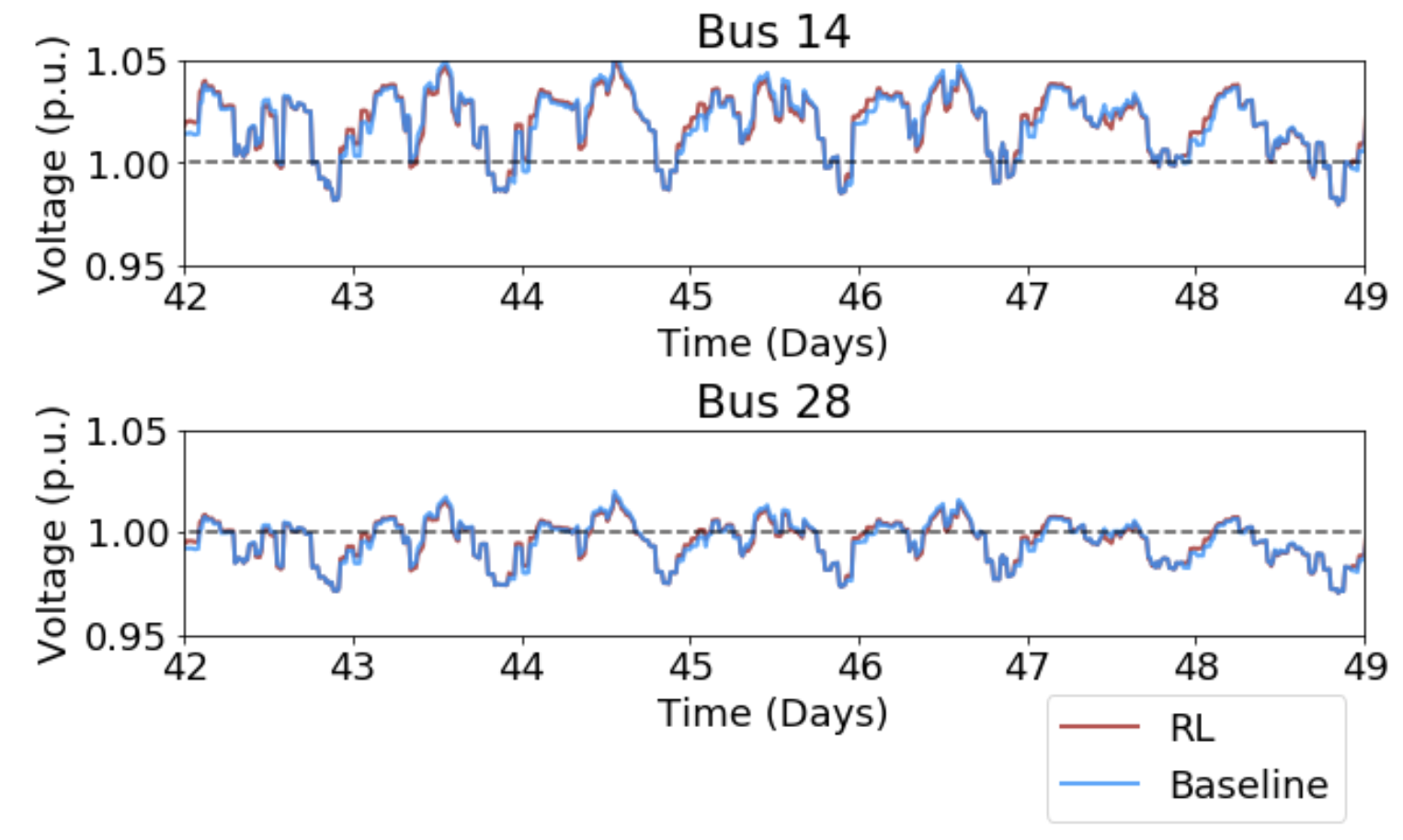}
    \caption{Voltages at Buses 14, 28}
    \label{fig:voltages}
\end{figure}

Once the RL agent is trained, it can be applied to different climate zones, or smaller time intervals. The results shown in this section are for models trained on Climate Zone 2A (hot, humid) and tested on Climate Zone 3A (warm, humid). Regardless of the change in climate data, the RL agents are successful in reducing the L2-norm of voltage deviations by 0.5\% on average. In Figure \ref{fig:voltages} voltages over 1 p.u. are reduced (with the largest deviations seeing the greatest reduction) and some voltages under 1 p.u. are raised. In the winter months, where the voltage is generally above 1 p.u., it seems that the RL agents becomes biased towards voltage reduction.

\begin{figure}
    \centering
    \includegraphics[width=\linewidth]{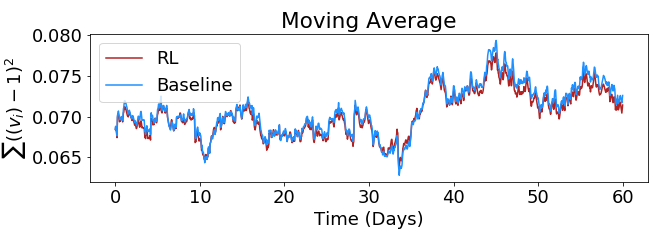}
    \caption{Overall grid voltage deviation}
    \label{fig:voltage_norm}
\end{figure}

\begin{figure}
    \centering
    \includegraphics[width=\linewidth]{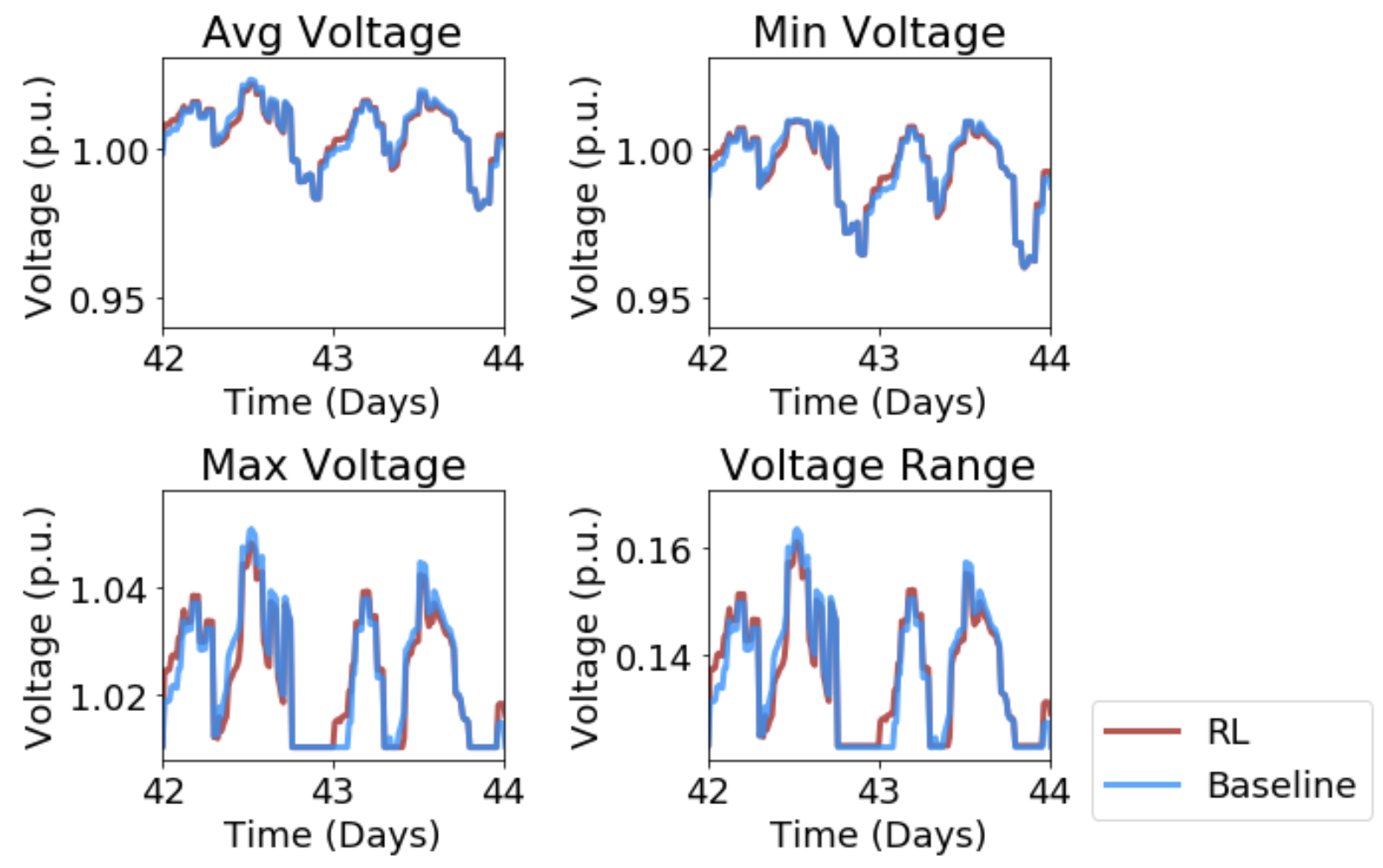}
    \caption{Network voltage summary}
    \label{fig:voltage_summary}
\end{figure}

\begin{figure}[t]
    \centering
    \includegraphics[width=\linewidth]{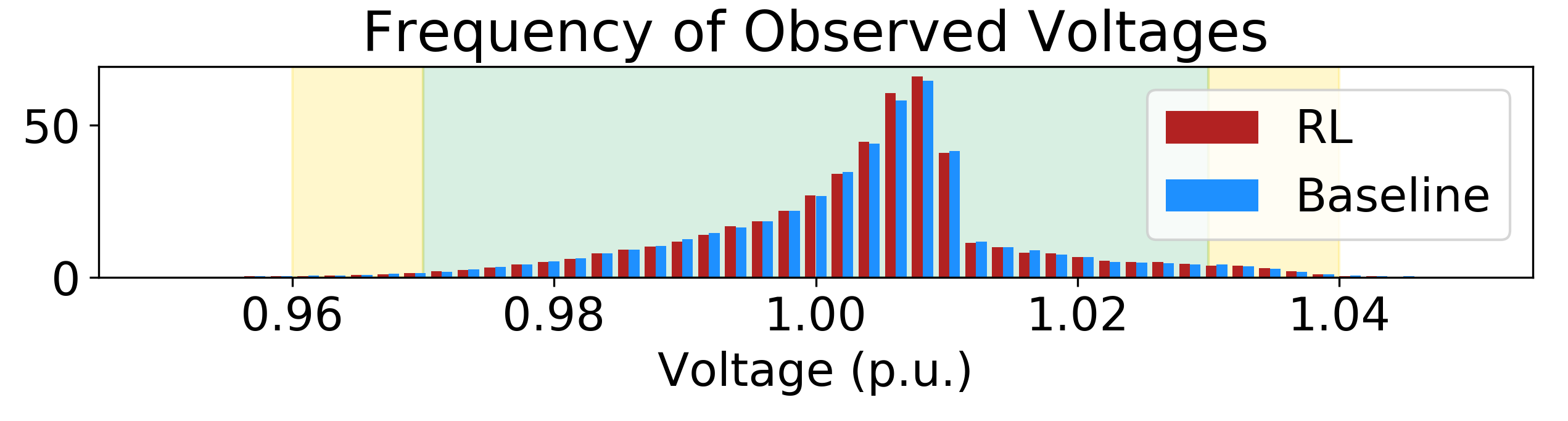}
    \caption{Histogram of observed voltages}
    \label{fig:voltage_hist}
\end{figure}

In Figure \ref{fig:voltages} the RL agents at times perform worse at Bus 14 or 28. For example, at the beginning of day 43, the voltage is raised for both Bus 14 and Bus 28. In Figure \ref{fig:voltage_summary} we can see that the maximum voltage is temporarily raised even further than the baseline due to this change in voltage. However, in Figure \ref{fig:voltage_hist} we also see that the average voltage hovers around 1 p.u. and the highest peak values of the maximum voltage are still reduced. This is likely due to the strong coupling between the buses where raising one voltage that is below 1 p.u. might also raise the voltage elsewhere in the network to be above 1 p.u. A moving average of the L2-Norm of the voltage deviation at all buses is plotted in Figure \ref{fig:voltage_norm}. This demonstrates that the multiagent approach still works to reduce voltage deviations in general.

Considering the overall voltage deviations (Fig \ref{fig:voltage_norm}), we observe that the RL agents generally reduce the overall voltage deviations, especially during periods of more extreme deviations (i.e. Days 40-60). In Figure \ref{fig:voltage_hist} we plot the number of intervals that each voltage is observed. In the baseline case, the number of 15-minute intervals across all 33 buses in which the voltage is at the upper bounds (1.03-1.05 p.u.) is more than twice the number of intervals that any bus sees a voltage at the lower bound (0.95-0.97 p.u.). The skew towards rising voltages aligns with earlier studies such as \cite{Dubey2017} and suggests that the most important task for the RL agents is in preventing overvoltages. The number of observed voltages over 1.03 and 1.04 p.u. as well as voltages under 0.97 and 0.96 p.u. are summarized in Table \ref{tab:heuristic_summary}. Note that the RL agent significantly reduces overvoltages that are closer to the operating bounds.

\begin{table}[t]
    \centering
    \begin{tabular}{c|c c c}
        & Baseline & RL & \% reduction \\
        $v_i^t > 1.04$ & 812 & 532 & 34.4 \\
        $v_i^t > 1.03$ & 6361 & 6156 & 3.2 \\
        $v_i^t < 0.97$ & 2867 & 2804 & 2.2 \\
        $v_i^t < 0.96$ & 1035 & 1018 & 1.6
    \end{tabular}
    \caption{Number of instances of over/under voltages}
    \label{tab:heuristic_summary}
\end{table}

\begin{figure}[]
    \centering
    \includegraphics[width=\linewidth]{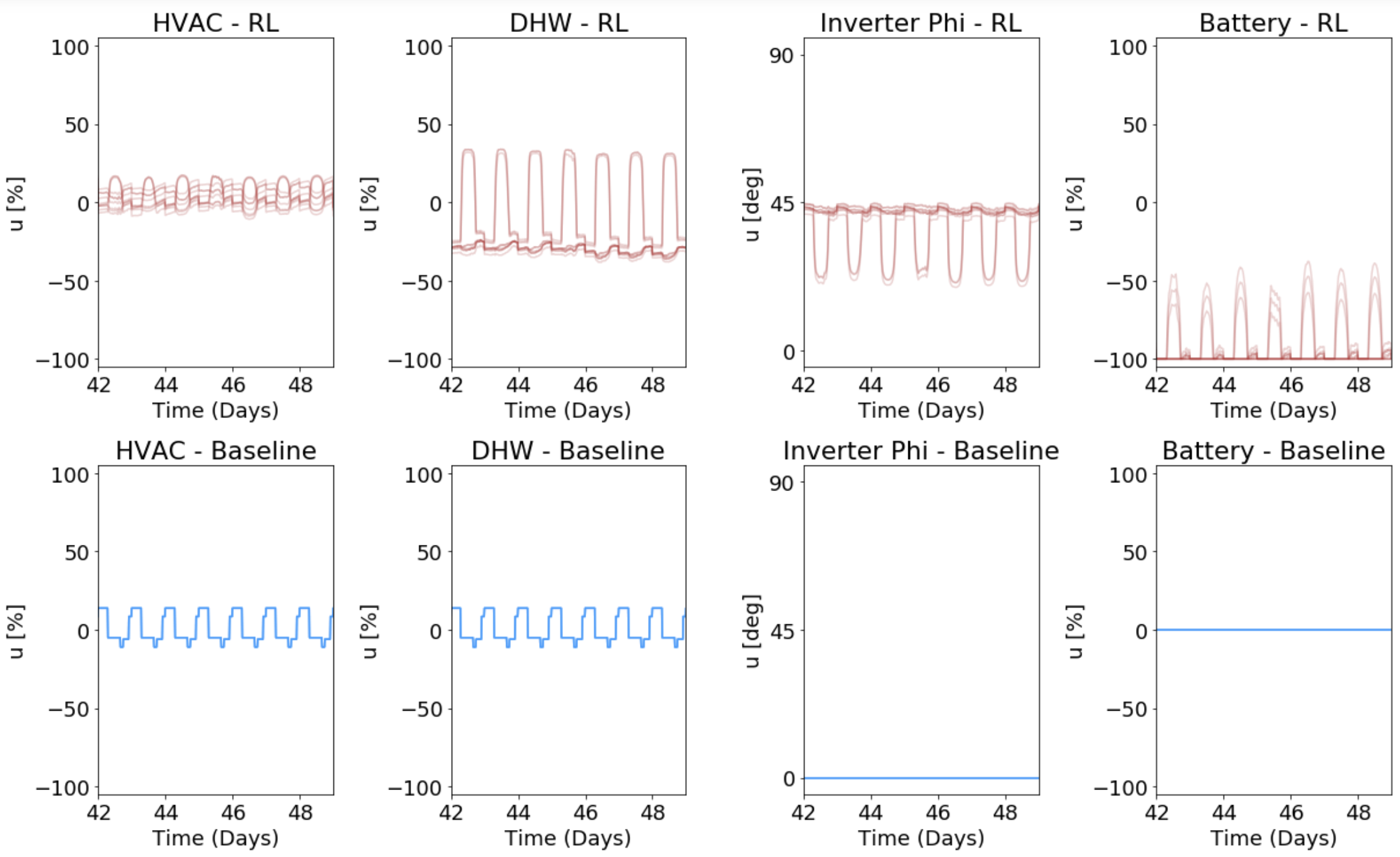}
    \caption{Subsystem action selection across buildings}
    \label{fig:actions}
\end{figure}

By inspecting the actions selected by each agent in Figure \ref{fig:actions}, we see that all agents converge on similar control strategies. This would likely lead to the RL agents learning to reduce their responsiveness, as they try to avoid collectively acting too responsively in over or undervoltage scenarios. In general, the RL agents learn to shift their peak demand to slightly later in the compared to the baseline RBC controllers, and learn to be more conservative in their charge/discharge signals. It should be noted that the battery subsystem completely stops charging and discharging during the testing period, even though it was helpful in training the agents. Since the battery is a DC resource behind the smart inverter, it likely strengthens the impact of the RL phase lag control. However, the battery itself might prove too responsive during training to be a legitimately useful resource in the testing phase. The increased reliance on thermal storage might also be beneficial to building owners since charge/discharge cycles add little wear to the HVAC and DHW systems and leaves the battery energy storage free for other resilience related objectives.

The inverter phase shift observed in Figure \ref{fig:actions} reveals that the RL agents learn to curtail real power injections in favor of reactive power injections that supplement the grid capacitor banks in providing voltage regulation. Grid operators might allow for increased real power uptake by increasing the size number of capacitor banks in the grid.

\begin{figure}[]
    \centering
    \includegraphics[width=\linewidth]{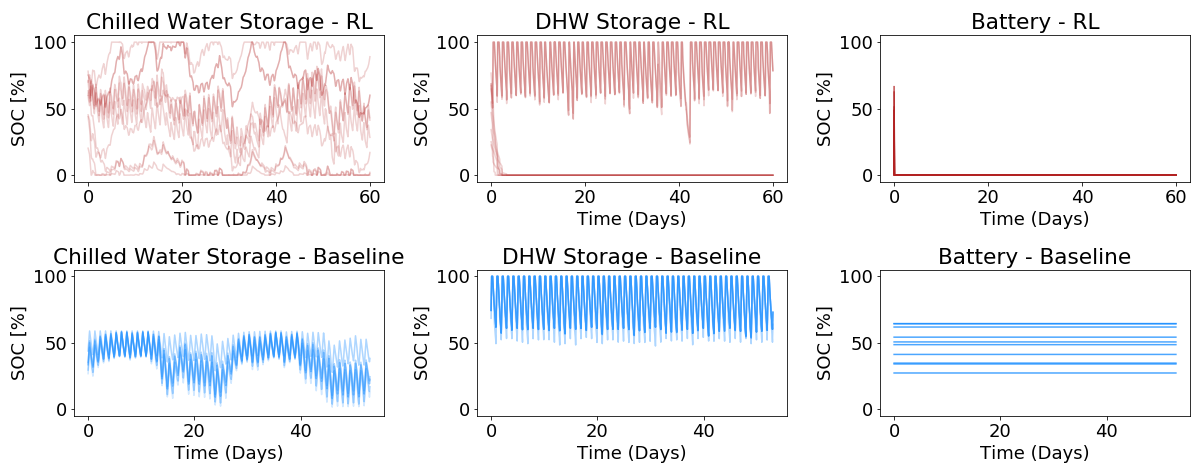}
    \caption{Subsystem state of charge across buildings}
    \label{fig:soc}
\end{figure}

The resulting state of charge for each of these scenarios is shown in Figure \ref{fig:soc}. As shown in Alg. \ref{alg:charge}, even though the battery is consistently in a state of ``discharging'' the actual power discharged from the device is 0 kW, as the battery is quickly drained at the start of the simulation period. To activate the battery as resource, we speculate that more specific objectives such as net energy reductions should be implemented in the reward function and weighted with voltage regulation.

\section{Conclusion}
\label{sec:conclusion}
In this paper we have shown the potential of multiagent reinforcement learning to achieve voltage regulation in a distribution network. Through the use of CityLearn's energy models and pandapower's power flow models, we have provided a district of buildings with thermal-electric models that measures voltage regulation (among other grid level metrics). The results show that decentralized control of smart inverters and energy storage can provide voltage regulation.

We believe GridLearn is a powerful tool for validating the impact of building energy management on distribution networks. Future work could build on the results provided here by considering more diverse energy resources with a more diverse building portfolio, or varied building subsystems. Other architectures, such as asynchronous updates to the building controllers should be explored to determine if learning could be improved with fewer agents acting simultaneously.

\bibliography{refs}
\bibliographystyle{ieeetr}

\end{document}